\documentclass[twocolumn,showpacs,nofootinbib,tightenlines,nobibnotes, aps, amssymb,amsmath,prl,superscriptaddress]{revtex4}
\usepackage{graphicx}% Include figure files
\usepackage{epsfig}% Include figure files
\usepackage{wasysym}

\newcommand{\la}{\langle}
\newcommand{\ra}{\rangle}

\newcommand{\nn}{\nonumber}

\newcommand{\be}{\begin{eqnarray}}
\newcommand{\ee}{\end{eqnarray}}

\begin{document}
%\preprint{APS/123-QED}
%draft
%\twocolumn[\hsize\textwidth\columnwidth\hsize\csname @twocolumnfalse\endcsname
\title{Unusual liquid state of hard-core bosons on the pyrochlore lattice}
\author{Argha Banerjee}
\affiliation{Department of Theoretical Physics,
Tata Institute of Fundamental Research,
1, Homi Bhabha Road, Colaba, Mumbai 400005, India}
\author{Sergei V. Isakov}
\affiliation{
Department of Physics, University of Toronto, Toronto, Ontario M5S 1A7, Canada}
\author{Kedar Damle}
\affiliation{Department of Theoretical Physics,
Tata Institute of Fundamental Research,
1, Homi Bhabha Road, Colaba, Mumbai 400005, India}
\author{Yong Baek Kim}
\affiliation{
Department of Physics, University of Toronto, Toronto, Ontario M5S 1A7, Canada}

\date{June 30, 2007}

\begin{abstract}
We study the physics of hard-core bosons with unfrustrated hopping ($t$) and nearest neighbour
repulsion ($V$) on the three dimensional pyrochlore lattice. At half-filling, we demonstrate that
the small $V/t$ superfluid state eventually becomes unstable at large enough $V/t$
to an unusual insulating state which displays no  broken lattice
translation symmetry. Equal time and static density
correlators in this insulator are well-described by a mapping to electric field
correlators
in the Coulomb phase of a $U(1)$ lattice gauge theory, allowing us to identify this insulator
with a $U(1)$ fractionalized
Mott insulating state. The possibility
of observing this phase in suitably designed atom-trap experiments with ultra-cold atoms is
also discussed, as are specific experimental signatures.
\end{abstract}

\vskip2pc

\maketitle

Much of our current understanding of the low temperature behaviour of condensed matter systems is based on
highly successful theoretical paradigms such as Landau's Fermi liquid theory of
normal metals, Bogoliubov theory for superfluids, BCS theory of superconductivity
and spin-wave theory for ferromagnets and antiferromagnets \cite{Andersonbook}. However, some systems exhibit behaviour that falls outside of any of these standard
paradigms---one example of this is the unconventional
normal state of underdoped high-T$_{\mathrm c}$ superconductors \cite{hightcreview1,hightcreview2}, while
other examples include the cooperative paramagnetic state of frustrated magnets \cite{Moessnerreview},
and the unusual phenomenology of heavy fermion compounds \cite{SCESreview}.
For instance, in the underdoped normal state of high-T$_c$ superconductors, some of the experimental
evidence is suggestive of the fact that the elementary quasiparticles excitations
are not spin-$1/2$ charge-$e$ holes, but spinless charge carriers propagating separately from
chargeless spin carriers \cite{hightcreview1,hightcreview2}.

This has motivated much of the recent effort aimed at providing theoretically consistent
descriptions of low temperature phases of  matter that would
display such spin-charge separation, or more generally, quasiparticle
fractionalization.  These developments~\cite{Aletreview}
allow one to conclude that such exotic behaviour is indeed possible, and go on to provide a description of quasiparticle fractionalization in terms of an effective field theory with gauge symmetry~\cite{Read-Sachdev,Wen}. In this approach, fractionalized quasiparticles emerge as the true low energy excitations in deconfined phases of a gauge theory (in which the emergent gauge force is not strong enough to bind the fractionalized quasiparticles into more conventional quanta),
and can be accompanied by additional gauge excitations that carry energy but no
spin or charge (such as 
the vortex excitation of a $Z_2$ gauge theory~\cite{Senthil}).

A closely related strand of activity has focused on the analysis of particular microscopic
models in order to understand whether they exhibit such exotic phases for specific
values of input parameters. This has led, for instance, to the construction of several different models~\cite{Senthil_Motrunich,Balents_Fisher_Girvin,Isakov_Kim_Paramekanti}
which exhibit so called $Z_2$ deconfined phases (the nomenclature refers to
the effective gauge theory that affords the most `natural' description of the low-energy physics).

One may now ask: Is there an experimental system which
would display one of these fractionalized phases for a definite range of control parameters?
A promising avenue in this regard is the physics of ultra-cold atoms in optical lattice potentials.
Recent work has demonstrated that a wide variety of phenomena
of interest to condensed matter physics can be studied by appropriately engineering systems
of ultra-cold atoms in optical potentials. For instance, it has been possible to provide a cold-atom
realization of the superfluid-insulator transition in a bosonic hubbard model
with on-site interactions on a cubic lattice~\cite{Jaksch,Greiner}.  This has been followed by
several interesting
proposals for realizing  fermionic and bosonic models with a variety of tunable interactions in
different optical lattice geometries~\cite{Duan,Santos}.

\begin{figure}[t]
\includegraphics[height=3cm]{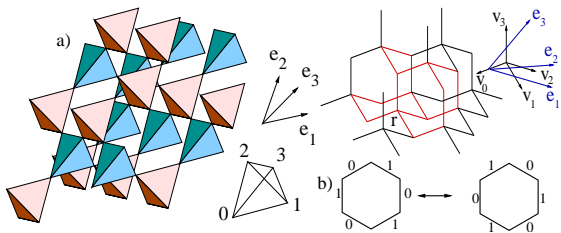}
\caption{(color online). (a) Pyrochlore lattice and the underlying diamond lattice. (b) Ring exchange process on plaquettes
of the diamond lattice.}
\label{lattice}
\end{figure}

In this work, we use sophisticated Quantum Monte Carlo (QMC) methods
to provide the first confirmation of the existence of
a $U(1)$ fractionalized insulating phase that may be realized in cold-atom
systems modeled by the Hamiltonian:
\be
H &=& \sum_ {\la ij\ra}[V(n_{i}-1/2)(n_{j}-1/2) -t(b^{\dagger}_{i}b_{j} + b_{i} b^{\dagger}_{j})] \nn \\
&&  +\sum_{i}[ U(n_i-1/2)^2 - \mu n_i] \; .
\label{eq:H_e}
\ee
Here, $n_i$ is the particle number
at sites $i$ of a three dimensional pyrochlore lattice (Fig~\ref{lattice} a), $b_i^{\dagger}$ is the corresponding boson creation operator, $U$ is the on-site repulsion and $V$ the nearest neighbour repulsion between bosons hopping (with amplitude $t$)
on  the nearest neighbour links $\la ij\ra$. 

Although the pyrochlore lattice geometry
we consider is technically challenging to realize, recent work that appeared as our study was underway
provides a viable prescription for experimentally realizing such an optical lattice ~\cite{Tewari}.
Furthermore, the simplicity of the interactions means that they can be
realized in state-of-the art cold atom experiments for a wide range of values of parameters~\cite{Jaksch}
including the `hard-core' limit ($n_i = 0, 1$) of very large $U$.
We therefore focus on this hard-core
limit in some detail here, setting $t=1$ and $\mu=0$ in what follows. [In this hard core limit, Eqn~\ref{eq:H_e} may also be written in spin $S=1/2$ language
via the mapping $S^z_i = n_i - 1/2$, $J_z = V$, $J_\perp = -2t$.] 

In this hard-core limit, with $\mu=0$ to enforce density $1/2$ per site, the physics at small $V$ is readily tractable: As the hopping
$t$ is {\em unfrustrated}, there is a stable superfluid phase at small $V$---indeed
a reasonable variational wavefunction for the ground state in this regime may be easily written down in spin language as $|\Psi \rangle =
\prod_i |S^x_i = +1/2 \rangle_i$.
What is the low temperature state in the opposite, large $V$ limit? To answer this, we
use the well-documented~\cite{Syljuasen0} stochastic
series expansion (SSE) QMC method (at large values of $V$, modifications developed recently~\cite{Gros} are crucial to
maintain ergodicity---for a review, see Ref.~[\onlinecite{Melko}]).

{\em Numerics: } Most of our results are on $L \times L \times L$ ($ L$, the number of up pointing
tetrahedra that fit in one side-length) samples with periodic boundary conditions and even $L$ ranging from $L=6$ to $L=12$, and inverse temperature $\beta$
ranging from $6$ to $120$ (with the largest $\beta$ employed for the largest size). We use standard SSE estimators\cite{Syljuasen0} to calculate the specific heat, the
superfluid stiffness $\rho_s$, the bond (kinetic) energy correlations, and the equal time $C^{\alpha \alpha^{'}}({\textbf{q}},\tau=0) = \langle n_\alpha({\textbf{q}}) n_{\alpha^{'}}(-{\textbf{q}})\rangle$ and static correlators $S^{\alpha \alpha^{'}}({\textbf{q}},\omega_n=0) = \int_{0}^{\beta}d \tau C^{\alpha \alpha^{'}}({\textbf{q}},\tau)$ of the density $n_i^{\alpha}$ (here $\alpha$, $\alpha^{'}$ refer
to different basis sites in a unit cell, and all site types (Fig.~\ref{lattice} (a)) are assigned coordinates of site-type $0$).

\begin{figure}[t]
\includegraphics[height=3cm]{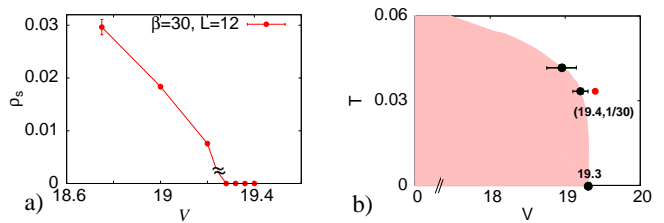}
\caption{(color online). (a) Superfluid density at $\beta=30$---the break around $V=
19.2$ indicates observed hysteresis near the
(weakly) first order transition. (b) Schematic phase diagram: black dots with error bars denote observed transitions, and
red dot denotes location at which
insulating phase data is displayed in Fig~{\protect{\ref{dipolar}}} and Fig~{\protect{\ref{highT}}}.}
\label{rho_ph}
\end{figure}

As is clear from Fig.~\ref{rho_ph} a), we see a distinct transition from a superfluid state at small $V$, to an insulating state
at large $V$ for a sequence of low temperatures. This transition is first-order at non-zero temperature
(Fig.~\ref{rho_ph} a), and while the first order nature is less prominent in lower temperature scans, a  scaling
analysis suggests that the transition remains first order even in the zero temperature limit~\cite{Isakov}. We estimate that this zero temperature transition is at $(V/t)_c \approx 19.3$ (Fig~\ref{rho_ph} b).

\begin{figure}[t]
\includegraphics[width=\hsize]{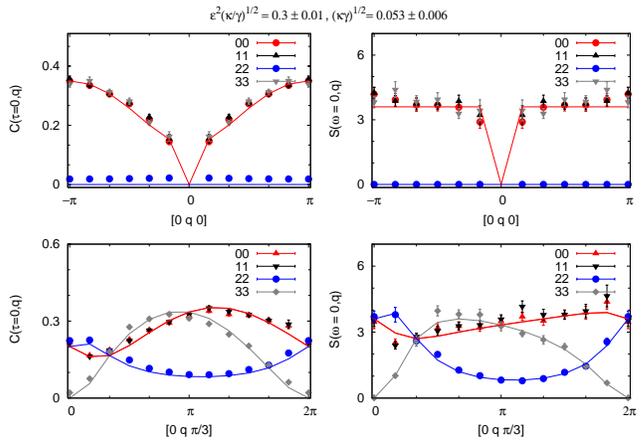}
\caption{ (color online). Static $S^{\alpha \alpha}({\bf{q}},\omega_n=0)$
and equal time $C^{\alpha \alpha}({\bf{q}}, \tau=0)$ correlators
of the density in the deconfined phase for $\beta =30$, $V=19.4$. The
lines are a fit to the predictions of non-compact $U(1)$ gauge theory on the diamond lattice
as discussed in text, with fit parameters displayed above.}
\label{dipolar}
\end{figure}

In the insulator, we see absolutely no Bragg peaks that would correspond to spatial
ordering in either the local density or the local bond energy. The insulator
is thus, in this specific sense, a liquid state of matter; this is illustrated in Fig~\ref{dipolar} with several scans of
density correlators in $q$ space at a representative point at very low temperature above the insulating ground state.
This absence of spatial ordering in the insulating state of an interacting boson system at $1/2$ filling is one of our striking results,
for such featureless insulating states are more typical of insulators with integer density per site.

{\em Interpretation:} Theoretical interpretation of this striking result is facilitated by noting that our Hamiltonian in this hard-core limit is closely related to
that studied in Ref.~\onlinecite{Hermele}:
Hermele {\em et al.} considered the $S=1/2$ XXZ antiferromagnet
on the pyrochlore lattice. By an analysis of a related
effective model of planar rotors (with additional terms added by hand to ensure
better
theoretical control), they argued that a $U(1)$
deconfined phase was a theoretically consistent possibility in the limit of extremely anisotropic exchange
$J_z \gg J_{\perp}> 0$---however, since the
positive sign of $J_\perp$ introduces a sign problem in quantum Monte-Carlo
treatment of such models, their work stopped short of making definitive statements about
the actual phase diagram of the $S=1/2$ model. [For a different
{\em effective} model of rotors that displays a $U(1)$ deconfined phase, see
Ref~\onlinecite{Lee}] 

Although our situation differs from Ref.~\onlinecite{Hermele} crucially in the
opposite
sign of the hopping term, this change of sign
does not affect \cite{Hermele} the arguments that make plausible the existence of a deconfined phase
at large $V/t$:  As the classical ground state for $t=0$ is macroscopically degenerate (with all configurations with two particles occupying
{\em each} tetrahedron having minimum energy), the
fate of the system at large but finite $V/t$ is then
determined by the structure of the effective Hamiltonian obtained to leading order in degenerate perturbation theory in $t/V$.\begin{figure}[t]
\includegraphics[width=\hsize]{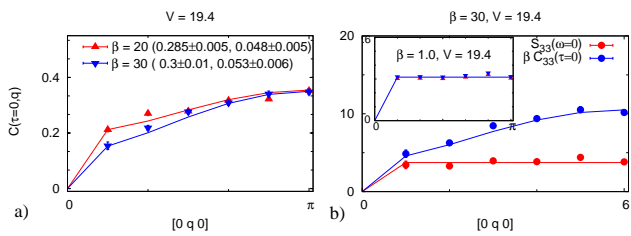}
\caption{ (color online). a) The same values of $\sqrt{\kappa \gamma}$ and $\epsilon^2 \sqrt{\kappa/\gamma}$ 
fit data at different temperatures for fixed $V/t$. b) $S^{\alpha \alpha}({\bf{q}},\omega_n=0)$
and $\beta C^{\alpha \alpha}({\bf{q}}, \tau=0)$ for $V=19.4$ are essentially equal at $\beta=1$,
but not at $\beta=30$.}
\label{highT}
\end{figure}

As the $t=0$ ground states can be represented in terms of dimers living on the links of the dual diamond lattice (Fig~\ref{lattice}(a)) subject to the constraint that two dimers touch each diamond lattice site, such a perturbative analysis yields
(at leading $O(t^3/V^2)$ order)  a quantum dimer model\cite{Hermele} with a `ring-exchange' term which causes a `flippable' hexagon (see Fig.~\ref{lattice} (b)) to resonate between its two allowed configurations.
The sign of this term is a matter of convention\cite{Hermele}
as it can be changed by an appropriate canonical transformation. However its structure, and
the structure of the constraint that defines the low energy manifold is highly reminiscent of
a (compact) $U(1)$ lattice gauge on the diamond lattice\cite{Hermele}. As the compact $U(1)$ theory
in three spatial dimensions admits a `Coulomb' phase that mimics ordinary electrodynamics,
we conclude, following Hermele {\em et al.}, that this is a consistent possibility at large
$V$ in our boson hubbard model.

{\em Fits:} To proceed further, we note that in such a deconfined phase, the low energy properties are expected to be described by the lattice version of standard Maxwell  electrodynamics with Hamiltonian ${\cal H}=\frac{{\gamma}}{2}\sum_{<\bf{r r'}>}e_{r r'}^2 + \frac{{\kappa}}{2} \sum_{\hexagon} (\Delta_{r r'} \times a_{r r'})^2$, where the lattice curl of the vector potential $a_{r r'}$ is defined on the hexagonal plaquettes of the dual diamond
lattice (Fig~\ref{lattice}(a)), the microscopic density operator is related to the
electric field by $n_{r r'} = \epsilon e_{r r'}$ with $\epsilon$ a non-universal scale factor, and ${\gamma}$ and ${\kappa}$
are the emergent energy scales of this low energy description. [Although $\gamma \sim U \rightarrow \infty$, $\kappa \sim
t^3/V^2 \rightarrow 0$ in the formal $V \rightarrow \infty$, $U \rightarrow \infty$ limit, their actual, renormalized values can be substantially different
from these bare estimates.]

To explore the implications of this ansatz for the density  correlators, it is useful to work with the corresponding
imaginary time action
\begin{eqnarray}
 S&=&\frac{1}{2}\stackrel{ \beta \sqrt{\kappa \gamma}}{\int_0} d\tilde{\tau} \left [ \sum_{\langle r r' \rangle}(\partial_{\tilde{\tau}} \tilde{a}_{r r'} -\Delta_{r r'} \tilde{a}_{\tau})^2 + \sum_{\hexagon} (\Delta_{r r'} \times \tilde{a}_{r r'})^2 \right ]  \; . \nonumber
\end{eqnarray}
Here, $\tilde{\tau} = \sqrt{\kappa \gamma} \tau$ is the dimensionless imaginary time variable obtained by scaling $\tau$ by
the typical photon energy $\sqrt{\kappa \gamma}$ of this artificial electrodynamics, $\tilde{a}_{r r'} = v^{1/4} a_{r r'}$ the rescaled
vector potential, and $\tilde{a}_{\tau} = v^{1/4} a_{\tau}/\sqrt{\kappa \gamma}$ the dimensionless scalar potential ($v= \kappa/\gamma$).
As density correlators are obtained by calculating corresponding  correlators of $\epsilon v^{1/4}(\partial_{\tilde{\tau}} \tilde{a}_{r r'} -\Delta_{r r'} \tilde{a}_{\tau})$ using this action, it is immediately clear that this electrodynamic ansatz predicts
 $C(\tau=0, {\textbf{q}}) = \epsilon^2\sqrt{v}f_{eq}(\beta \sqrt{\kappa \gamma},{\textbf{q}})$ and $S(\omega=0,{\textbf{q}}) =  \epsilon^2 \frac{\sqrt{v}}{ \sqrt{\kappa \gamma}}f_{st}({\textbf{q}})$.
In order to test this ansatz, we have calculated the functions $f_{eq}(\beta \sqrt{\kappa \gamma},{\textbf{q}})$
and $f_{st}({\textbf{q}})$  and performed detailed fits of our data for
the density correlators $C(\tau=0, {\textbf{q}})$ and $S(\omega=0,{\textbf{q}})$.

Our fitting procedure is quite straightforward: We first determine the best fit value
of the scale factor $c_{st}$ by which the function $f_{st}$ needs to be scaled to fit the static correlators $S$. Next, we
note that  the shape of $f_{eq}$ (as a function of ${\textbf{q}}$) depends significantly on the value of the typical photon energy  $\sqrt{\kappa \gamma}$ that enters its first argument, and
determine its best fit value such that $f_{eq}(\beta\sqrt{\kappa \gamma},\bf{q})$ best
mimics the {\em shape} of the corresponding equal time correlators $C_{eq}(\bf{q})$. Finally, we determine the best fit value of the corresponding equal time scale factor $c_{eq}$ by which the function $f_{eq}$ needs to be scaled to fit the overall magnitude of the equal time correlators $C_{eq}$.

Clearly, this is a very over-determined fit, since the {\em same} set of parameters have to fit scans of the correlators in the entire
brillouin zone, in addition to fitting data at different temperatures (at fixed $V/t$). In addition, this procedure has an in-built consistency check, since the value of photon energy scale
$\sqrt{\kappa  \gamma}$
can be re-obtained from the scale factors by noting that $c_{eq}/c_{st} = \sqrt{\kappa \gamma}$.

In Fig.~\ref{dipolar}, we show the results of such a  fit of the static (zero frequency) and equal time density correlators along
several scans in the Brillouin zone for a representative low temperature point  at which $\rho_s = 0$
(similar fits work equally well at other low temperature points in the insulating phase).  Clearly the data fits the predictions of non-compact electrodynamics
{\em extremely well}, with the best fit values of the photon energy scale $\sqrt{\kappa \gamma}$ and  $\epsilon^2 \sqrt{\kappa/\gamma}$ shown in Fig.~\ref{dipolar} (the quoted uncertainty in
the best fit value of $\sqrt{\kappa \gamma}$ also takes into account the accuracy with which the self-consistency condition is satisfied). Furthermore, for fixed $V/t$, the {\em same} parameters do indeed continue to fit the data as the temperature is varied (Fig~\ref{highT} a)).

{\em Discussion:} These fits are extremely convincing evidence that we have accessed the low temperature regime just above a Coulomb
liquid ground state. Nevertheless, it is instructive to play devil's advocate and ask if the measured correlators can satisfy the predictions of non-compact electrodynamics to this level of accuracy if the system does not have a deconfined coulomb liquid ground state? Perhaps surprisingly, the answer is yes, but only if the data has been taken at temperatures such that thermal fluctuations
overwhelm all quantum dynamics.

Assume for instance that the insulator is a more conventional lattice-symmetry breaking crystal that one may have expected at fractional filling. If one happens to be at a temperature above the melting temperature of this putative crystal, then thermal fluctuations would completely overwhelm quantum effects, and the physics would be essentially classical.
As long as the temperature remains much smaller than $V$, this classical physics  is correctly described by the classical dimer
model on the diamond lattice, regardless of the quantum ground state.

Now, static and equal time correlators of any quantum system are proportional
to each other (with proportionality constant $\beta$) in any such effectively classical regime.
Thus, we expect dimer correlations of the classical
dimer model to correctly describe the functional form of {\em both} the static {\em and} equal time correlators of the system in this regime.
Furthermore, these classical dimer correlators are known~\cite{Huse} to have precisely the same functional
form as the static correlators of ${\cal H}$.

Regardless of the quantum ground state, we thus expect our data for static {\em and} equal time correlation functions to be
{\em necessarily proportional} to each other {\em and} match predictions of
quantum electrodynamics in this classical regime.
Is this `trivial' mechanism responsible for the extremely good fits shown in Fig~\ref{dipolar}?  The answer is clearly {\em no}: If this were
the case, the static and equal time correlators, being proportional to each other, would have the same shape
(as a function of $\bf{q}$). This is {\em clearly not} the case for the low temperature data shown in Fig~\ref{dipolar}, as is
underscored by a comparison to data at {\em much higher temperatures} (Fig~\ref{highT} b)), where
this commonality of shape {\em does} become clearly visible.

The weight of all  this  evidence thus allows us to conclude that we are indeed seeing a Coulomb liquid state of matter in
our simulations. What would be the best way to `look' for this
state of matter in a putative cold-atom experiment? At the most gross level, this
phase is an incompressible insulator, with a gap to charged
excitations.
The distinctive difference from ordinary Mott insulating phases
(such as those seen in the experiments of Ref.~\onlinecite{Greiner}) is the presence of a gapless neutral collective mode, namely
the artificial photon of the $U(1)$ gauge theory mentioned above. 
As we have demonstrated above, this neutral mode leads to  characteristic dipolar structure
in the low temperature equal time and static density correlators. These correlations can
be measured in atom-trap experiments by noise correlation~\cite{Altman} measurements that
probe equal
time correlators, and Bragg scattering experiments~\cite{Stenger_etal} that probe static correlators.

{\em Acknowledgements:} We acknowledge useful discussions with L.~Balents, R.~Moessner, S.~Minwalla, A.~Paramekanti, and
T.~Senthil, computational resources
of TIFR, and support from
NSERC, CRC, CIAR, KRF-2005-070-C00044.(YBK \& SVI), and DST-SR/S2/RJN-25/2006 (KD).


\begin{thebibliography}{999}

\bibitem{Andersonbook} {\it Basic Notions of Condensed matter physics}, P.~W.~Anderson,
Addison-Wesley Publishing Company (1984).

\bibitem{hightcreview1} A.~Damascelli, Z.~X.~Shen,  and Z.~Hussain,  Rev. Mod. Phys. {\bf 75}, 473 (2003).

\bibitem{hightcreview2} P. A. Lee, N. Nagaosa, and X.-G. Wen, Rev. Mod. Phys. {\bf 78}, 17-85 (2006).

\bibitem{Moessnerreview} R.~Moessner, Can. J. Phys. {\bf 79}, 1283 (2001).

\bibitem{SCESreview} P.~Coleman, Physica B: Condensed Matter {\bf 378-380}, 1160 (2006).


\bibitem{Aletreview} F.~Alet, A.~M.~Walczak, and M.~P.~A.~Fisher, Physica A {\bf 369}, 122 (2006).

\bibitem{Read-Sachdev} 
N.~Read and S.~Sachdev, Phys. Rev. Lett. {\bf 66}, 1773 (1991).
\bibitem{Wen} 
X.~G.~Wen, Phys. Rev. B {\bf 44}, 2664 (1991).
 
\bibitem{Senthil} T. Senthil and M. P. A. Fisher, Phys. Rev. B {\bf 62}, 7850 (2000).

\bibitem{Senthil_Motrunich} O.~I.~Motrunich and T.~Senthil, Phys. Rev. Lett. {\bf 89}, 277004 (2002); T.~Senthil
and O.~I.~Motrunich, Phys. Rev. B {\bf 66}, 205104 (2002).

\bibitem{Balents_Fisher_Girvin} L.~Balents, M.~P.~A. Fisher, and S.~M.~Girvin
Phys. Rev. B {\bf 65}, 224412 (2002). 

\bibitem{Isakov_Kim_Paramekanti}S. V. Isakov, Y. B. Kim, and A. Paramekanti, Phys. Rev. Lett. {\bf  97}, 207204 (2006).


\bibitem{Greiner} M.~Greiner {\it et al.}, Nature {\bf 415} 39 (2002).

\bibitem{Jaksch} D.~Jaksch {\it et al.}, Phys. Rev. Lett. {\bf 81}, 3108 (1998).
 




\bibitem{Duan} L.~M.~Duan, E.~Demler, and M.~Lukin, Phys. Rev. Lett. {\bf 91},
090402 (2003).

\bibitem{Santos} L.~Santos {\it et al.}, Phys. Rev. Lett. {\bf  93}, 030601 (2004).


\bibitem{Tewari} S.~Tewari {\it et al.}, Phys. Rev. Lett. {\bf 97}, 200401 (2006).

\bibitem{Syljuasen0} O.~Syljuasen and A.~W.~Sandvik, Phys. Rev. E {\bf 66}, 046701
(2002).


\bibitem{Gros} K.~Louis and C.~Gros, Phys. Rev. B {\bf 70}, 100410(R) (2004);
K.~Damle {\em et al.}, unpublished.

\bibitem{Melko}  R.~G.~Melko, J. Phys.: Condens. Matter \textbf{19}, 145203 (2007).

\bibitem{Isakov} S.~V.~Isakov, A.~Banerjee, K.~Damle, and Y.~B.~Kim, unpublished.


\bibitem{Hermele} M.~Hermele, M.~P.~A.~Fisher, and L.~Balents,
Phys. Rev. B {\bf 69}, 064404 (2004).

\bibitem{Lee} S.~Lee and P.~A.~Lee, Phys. Rev. B {\bf 74}, 035107 (2006).




\bibitem{Huse} D.~A.~Huse, W.~Krauth, R.~Moessner, and S.~L.~Sondhi,
Phys. Rev. Lett. {\bf 91}, 167004 (2003).
 



\bibitem{Altman} E.~Altman, E.~Demler, and M.~D.~Lukin, Phys. Rev. A {\bf 70}, 013603 (2004).

\bibitem{Stenger_etal} J.~Stegner {\it et. al.},  Phys. Rev. Lett. {\bf  82}, 4569 (1999); D.~M.~Stamper-Kurn {\it et. al.} Phys. Rev. Lett. {\bf  83}, 2876 (1999).










\end{thebibliography}
\end{document}